\documentclass[3p,times,procedia]{elsarticle}
\usepackage{ecrc}
\volume{00}

\firstpage{1}

\journalname{Energy Procedia}

\runauth{Wlaz{\l}o et al.}

\jid{procs}

\usepackage{amssymb}
\usepackage[utf8]{inputenc}
\usepackage{siunitx}
\usepackage{tabularx}
\usepackage{cleveref}

\begin{document}
	
\begin{frontmatter}



\dochead{European Geosciences Union General Assembly 2017, EGU\\ Division Energy, Resources \& Environment, ERE}%
%% e.g. \dochead{17th International Conference on Dynamical Processes in Excited States of Solids}

\title{Ab initio studies of carbon dioxide affinity to carbon compounds and minerals}

\author[a]{Mateusz Wlazło\corref{cor1}} 
\author[a,b]{Alexandra Siklitskaya}
\author[a]{Jacek A. Majewski}

\address[a]{Faculty of Physics, University of Warsaw, Pasteura 5 02-093 Warsaw, Poland}
\address[b]{Institute of Physical Chemistry, Polish Academy of Sciences, Kasprzaka 44/52, Warsaw, Poland}

\begin{abstract}
We have performed quantum chemical computational studies to determine carbon dioxide affinity to carbon compounds and minerals, which could be present in shales. These studies shed light on the microscopic mechanisms of the possible carbon oxide sequestration processes. Our studies reveal that the carbon oxide can be adsorbed to various forms of carbon structures and also minerals such as periclase or illite. We find out that the strongest affinity of carbon oxide towards carbon structures occurs when the carbon structures exhibit $sp^3$ bonds.
\end{abstract}

\cortext[cor1]{Corresponding author}
\end{frontmatter}

%\correspondingauthor[*]{Corresponding author. Tel.: +0-000-000-0000 ; fax: +0-000-000-0000.}
\email{mateusz.wlazlo@fuw.edu.pl}

% \linenumbers


%\enlargethispage{-7mm}
\section{Introduction}

% In this work we employ quantum chemical computational techniques to study carbon dioxide adsorption processes on the atomistic level in organic and inorganic shale rock constituents. We have selected a few surfaces that mirror the heterogeneity of shales. In particular, we focus on the organic components, which are modeled by pure carbon allotropes.

% Total organic carbon (TOC) is an important parameter that characterizes the potency of a shale formation in hydrocarbon production. A high TOC value ensures that a reservoir is useful in gas production.

Carbon dioxide is one of the primary greenhouse gases implicated in global warming process. Alternative energy sources that make no contribution to $CO_2$ emission are still in the development phase and not likely to replace current carbon-based energy sources during the few next decades. Also carbon capture methods are intensively studied at present, however, it is difficult to foreseen their effective employment in industry and everyday life. Carbon dioxide sequestration process seems to be an alternative, and therefore, is of critical importance to maintain or even reduce the $CO_2$ level in the atmosphere. Additionally, $CO_2$ sequestration process can be connected to the Enhanced Oil recovery in shale rocks. Therefore, accurate description of physicochemical mechanisms governing the $CO_2$ and $CH_4$ adsorption to various minerals and organic forms of carbon matter is an important step to understand mechanisms and to improve technologies of sequestration. The particular role of carbon in these processes is evident, since total organic carbon (TOC) is an important parameter that characterizes the potency of a shale formation in hydrocarbon production, and a high TOC value ensures that a reservoir is useful in gas production.

In the present study, we employ quantum chemical computational techniques to study carbon dioxide adsorption processes on the atomistic level in organic and inorganic shale rock constituents and to gain understanding of these effects on the micro (atomic) scale. We have selected a few materials that mirror the heterogeneity of shales. In particular, we focus on the organic components, which are modeled by pure carbon allotropes such as graphene and spiral carbon nanoparticles (spiroids), however, we provide also some findings for magnesium oxide mineral and illite rocks. We investigate the structural, energetic, and the thermodynamical aspects of $CO_2$ adsorption and desorption employing ab initio molecular dynamics (AIMD) within the canonical ensemble (NVT) and are able to determine the possible $CO_2$ capture reactions. 

The present paper is organized as follows. In \Cref{section:Methods} we give a short and comprehensive description of the methodology employed in this study. The results of the study describing the mechanisms of the $CO_2$ adsorption to the carbon structures and minerals are presented in \Cref{section:Results}. Finally, the paper is concluded in \Cref{section:Conclusions}. 

\section{Methods}
\label{section:Methods}
\begin{figure}
	\centering
	\includegraphics[width=.6\linewidth,height=.4\linewidth,keepaspectratio]{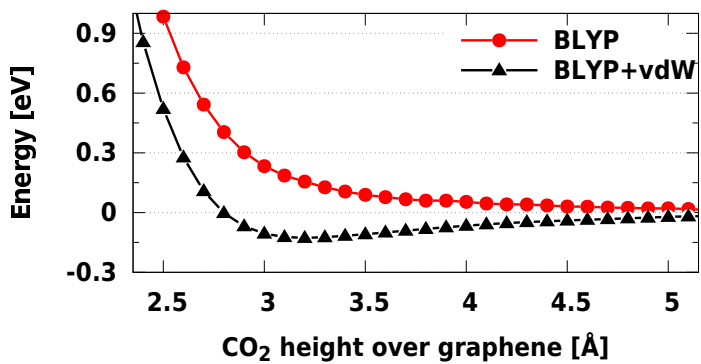}
	\caption{Adsorption energy vs. distance of $CO_2$ over ideal graphene obtained in calculations with (BLYP+vdW) and without (BLYP) the DFT-D2 correction. The presence of an energy minimum after including the correction indicates that $CO_2$ can be physisorbed on graphene and that it can only occur via van der Waals interaction. Reprinted from \cite{APPAMW}.}
	\label{figure:vdw}
\end{figure}

% We employ density functional theory (DFT) calculations to find the ground states of analyzed systems. The Hohenberg-Kohn-Sham \cite{DFT-HK,DFT-KS} realization of DFT allows us to find every observable of the system as a functional of the electronic ground state. The sum of these observables is the total energy functional, often referred to as the Kohn-Sham functional ($E_{KS}$) that is written in terms of the kinetic energy functional ($T_s$), the external potential acting on the electronic system ($v_{ext}$), the electron-electron interaction (Coulomb or Hartree interaction, $E_H$) and exchange-correlation interaction ($E_{XC}$):

The essential component of {\it ab initio} methods for predicting properties of a collection of atoms is a quantum mechanical scheme to calculate the ground state energy of the system studied. In our studies of the physicochemistry of the carbon dioxide adsorption to surfaces of various materials, we employ density functional theory (DFT) calculations that constitute nowadays the method of choice for the materials science. The Hohenberg-Kohn-Sham \cite{DFT-HK,DFT-KS} realization of DFT allows us to find not only the total energy of the system but also every observable of the system as a functional of the ground state electronic density. The total energy functional, often referred to as the Kohn-Sham functional ($E_{KS}$), is written in terms of the kinetic energy functional ($T_s$), the external potential acting on the electronic system ($v_{ext}$), the electron-electron interaction (Coulomb or Hartree interaction, $E_H$) and exchange-correlation interaction ($E_{XC}$):

\begin{equation}
	E_{KS}[n] = T_s[n] + \int d^3r\ v_{ext}({\bf r}) n({\bf r}) + U_H[n] + E_{XC}[n]
\end{equation}

The exact form of the XC term is unfortunately unknown and it has to be approximated \cite{Martin}. The choice of a XC functional is a crucial parameter of a DFT calculation. Many successful approaches have been based on the local density approximation (LDA), which is derived from the homogeneous electron gas model. However, the so-called gradient-corrected functionals, or generalized gradient approximations (GGA), are superior to LDA. Therefore, we use the GGA form of the exchange correlation functional, as implemented in the so-called BLYP (Becke-Lee-Yang-Parr) form \cite{Becke,LYP}. Unfortunately, the GGA approach does not include dispersion interaction, which is a necessity if one wishes to account for van der Waals interaction. For this reason, we add another term to the KS functional. It is a semi-empirical, pairwise D2 correction \cite{DFT-D2}, which has the following form:

\begin{equation}
	E_{D2} = -s_6 \sum_{A<B} \frac{C_6^{AB}}{R_{AB}^6}f_{damp}(R_{AB},R_{0AB}),
\end{equation}
where the sum runs over pairs of atoms A and B, $R_{AB}$ is the interatomic distance, $C^6_{AB}$ are parameters obtained from time-dependent DFT \cite{DFT-D2}, $R_{0AB}$ is the sum of van der Waals radii of atoms A and B, $s_6$ is a fitted parameter. The full functional therefore reads:

\begin{equation}
	E_{DFT-D2} = E_{KS}[n] + E_{D2}.
\end{equation}

Including the D2 correction is necessary to obtain correct physisorption energy profiles, as it is demonstrated for the case of adsorption of $CO_2$ to the graphene surface in \Cref{figure:vdw}. As it is seen, the adsorption energy as a function of the distance of the $CO_2$ molecule to the graphene surface has no minimum when the BLYP functional without van der Waals corrections is employed for exchange correlation functional, indicating that the $CO_2$ does not adsorb to the surface. Only inclusion of van der Waals forces to the energy functional guarantees physisorption of $CO_2$ to the graphene layer at the distance corresponding to the minimum of the BLYP+vdW total energy curve.

Further in the text we will refer to adsorption or interaction energy. These quantities are calculated using the total energy functionals. For instance, we calculate the adsorption energy of a species on a surface by calculating the total energy of a system containing both the species and the surface. Then we subtract the sum of energies of systems containing only the surface and only the species, i.e.:

\begin{equation}
	E_{ads} = E_{surf.+species} - (E_{surf.}+E_{species}).
\end{equation}
Adsorption is energetically favored whenever $E_{ads} < 0$.

Once the ground state Hamiltonian has been found from DFT, forces acting on nuclei (i.e., the derivatives of the total energy with respect to the atomic positions) can be calculated according to the Hellmann-Feynman theorem \cite{Hellmann-Feynman}.

A widely used DFT geometry optimization scheme involves minimizing Hellmann-Feynman forces. For an initial geometry, forces are calculated from the ground state and then the atoms are moved in subsequent steps until an arrangement is found for which the forces vanish. \Cref{figure:sw_graphene,figure:spiroid,figure:minerals} show geometries which are the result of such optimizations.

In {\it ab initio} molecular dynamics (AIMD) simulations, interatomic forces are calculated according to this relation and the motion of nuclei is propagated with Newton's equations of motion. There are several different schemes to treat the electronic system in such a simulation. Here we use the Car-Parrinello approach \cite{Car-Parrinello}. This scheme is extended by coupling to external Nose-Hoover thermostats \cite{nosethermostat,hooverthermostat}. This allows for simulations in the canonical (NVT) ensemble.

\section{Results}
\label{section:Results}
\subsection{$CO_2$ adsorption mechanisms}
	There are several chemical mechanisms of adsorption of and $CO_2$ on surfaces, including direct chemisorption, dissociative chemisorption and physisorption. Depending on many factors, such as electronic structure, molecular and surface charge distribution or geometry, different mechanisms will dominate the adsorption processes. For example, the methane molecule $CH_4$ has a closed shell electronic configuration. All of its valence electrons form bonds within the molecule. Therefore, direct chemisorption of such species can be ruled out immediately. An intermediate dissociation step has to occur first, i.e.:

	\begin{equation}
		CH_4(gas) \rightarrow CH_3(ads) + H(ads)
	\end{equation}

	Even if the dissociation reaction is endothermic, there is an energy barrier for it to occur. The magnitude of the barrier depends on the adsorbent. In many cases physisorption of $CO_2$ is favored. It is much weaker than chemisorption but it plays a significant role in gas reactions in interporous space.

	Another example is the direct chemisorption of $CO_2$. Under normal conditions, the molecule is linear with two double bonds between C and O atoms. Chemisorption requires breaking one of the C-O bonds which greatly distorts the shape of the molecule. Such a distortion is associated with a high energy barrier. Typical reservoir conditions, however, are well above the supercritical point for $CO_2$. First principles calculations show slight non-linearity of supercritical $CO_2$ \cite{scCO2}. This means that the activation energy for direct $CO_2$ chemisorption is decreased in the supercritical fluid phase.
	
	\subsection{Adsorption on different classes of materials}
	\begin{table}[b]
	\caption{Dominant chemisorption mechanisms and $CO_2$ adsorption energies for analyzed structures. See \Cref{figure:sw_graphene,figure:spiroid,figure:minerals} for optimized geometries.}
	{\setlength{\extrarowheight}{10pt}%
	\begin{tabularx}{\linewidth}{l|XXXXX}
	\hline
	Species & SW-defected graphene & $C_{300}$ spiroid & Calcite ($CaCO_3$) & Periclase (MgO) & Illite ($Al_4KSi_2O_{12}$)\\
	Dominant chemisorption mechanism & Direct & Direct & Dissociative & Direct & Direct\\
	Adsorption energy [\si{\kJ\per\mole}] & -143 & -15 & -98 & -90 & -92\\
	\hline
	\end{tabularx}}
	\end{table}

	In this section we will describe adsorption mechanisms on a variety of carbon allotropes. As the first one we consider graphene, which is an atomically-thin, planar structure with hexagonal arrangement of atoms in a honeycomb-like lattice. Layers of graphene stacked on top of each other form graphite. It can also serve as a model of dehydrogenated kerogen. We will also consider two spherical forms of carbon -- multishell fullerene and a carbon spiroid.

		\subsubsection{Graphene}
			\begin{figure}
			\centering
			\includegraphics[width=.45\linewidth,height=.3\linewidth,keepaspectratio]{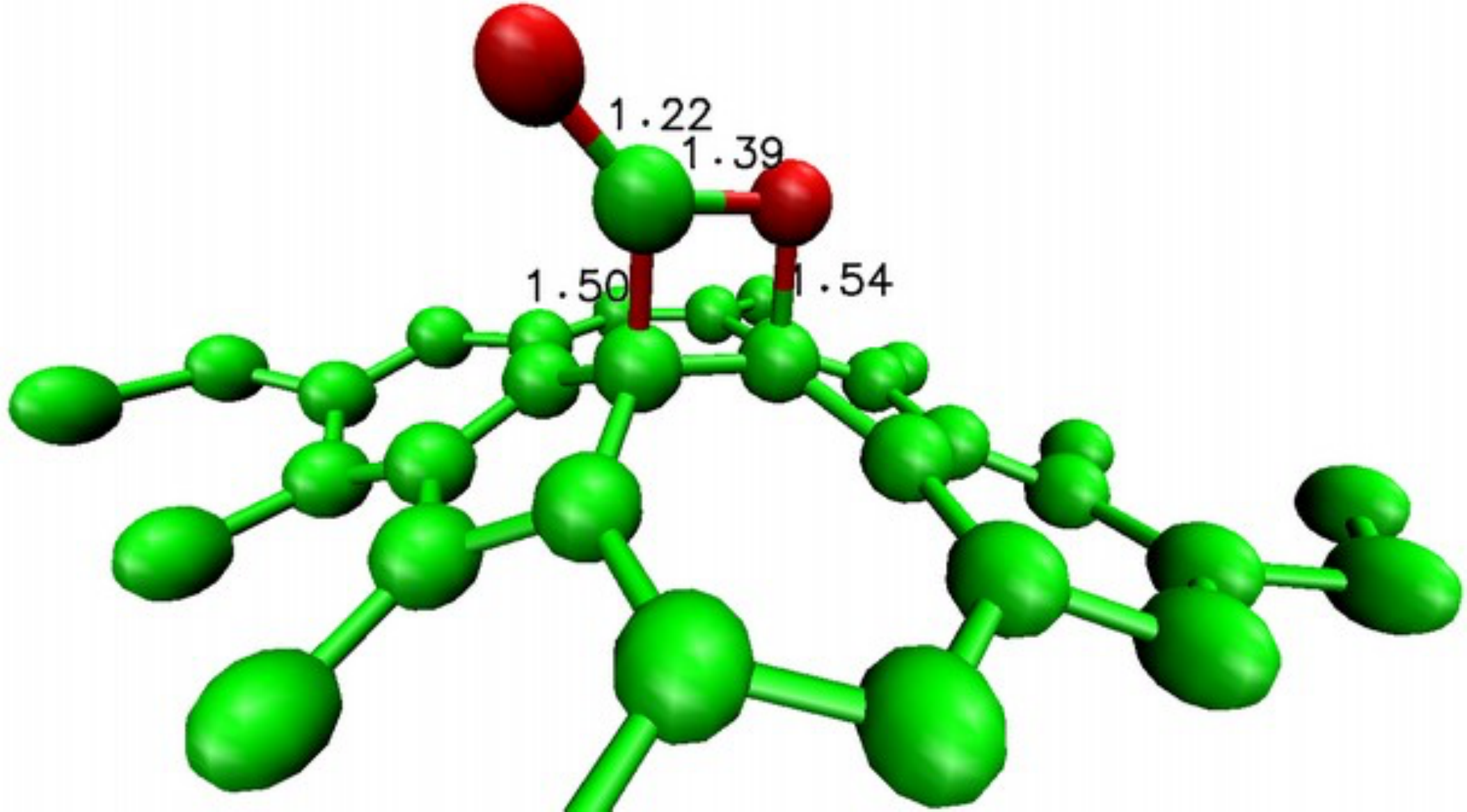}
			\caption{Optimized geometry of $CO_2$ chemisorbed on a Stone-Wales defect site in graphene.}
			\label{figure:sw_graphene}
			\end{figure}

			Graphene lattice is made up entirely of $sp_2$ carbon atoms. Atoms are bonded by in-plane $\sigma$ bonds and delocalized $\pi$ bonds formed by type p orbitals that are perpendicular to the surface. When another molecule or radical approaches the surface, it is possible for the $\pi$ bond to be disturbed. If the species has a free electron of its own, the dangling $p_z$ orbital in graphene can overlap with the unoccupied orbital of the species. Then another $\sigma$ bond is made and the $sp_2$ carbon undergoes rehybridization to $sp_3$. Because $sp_3$ carbon geometry is tetrahedral (as in the $CH_4$ molecule) rather than planar (as in graphene), the atom moves out of plane by about 0.35 \si{\angstrom}.

			Graphene lattice can feature a plethora of structural defects. The simplest ones are so-called point defects that only change a single atom, e.g. a lattice vacancy or single atom substitution by another atom type. A different type of defect is the Stone-Wales (SW) defect which is created by rotating one C-C bond by 90 degrees \cite{SW-CPL}. In the neighborhood of such a defect the hexagonal lattice is disturbed an instead of hexagonal rings we obtain two heptagons and two pentagons. This defect does not change the $\sigma$ bonds -- every carbon atom still has three neighbors. Local stress that appears in this structure is released by buckling around the defect \cite{SW-buckling}. This in turn disturbs the $\pi$-bond system and makes the site more reactive towards chemisorption \cite{APPAMW}

			The chemistry of SW-defected lattice is slightly different. This has an important implication when it comes to $CO_2$ adsorption. On ideal graphene lattice, there is no stable chemisorption of $CO_2$ and there is only weak physisorption. From geometry optimization of $CO_2$ on SW defect we find a stable configuration in which $CO_2$ is chemisorbed by two covalent C-C and C-O bonds (see \Cref{figure:sw_graphene}).

		\subsection{Spherical forms of carbon}
			\begin{figure}
			\centering
			\includegraphics[width=.45\linewidth,height=.3\linewidth,keepaspectratio]{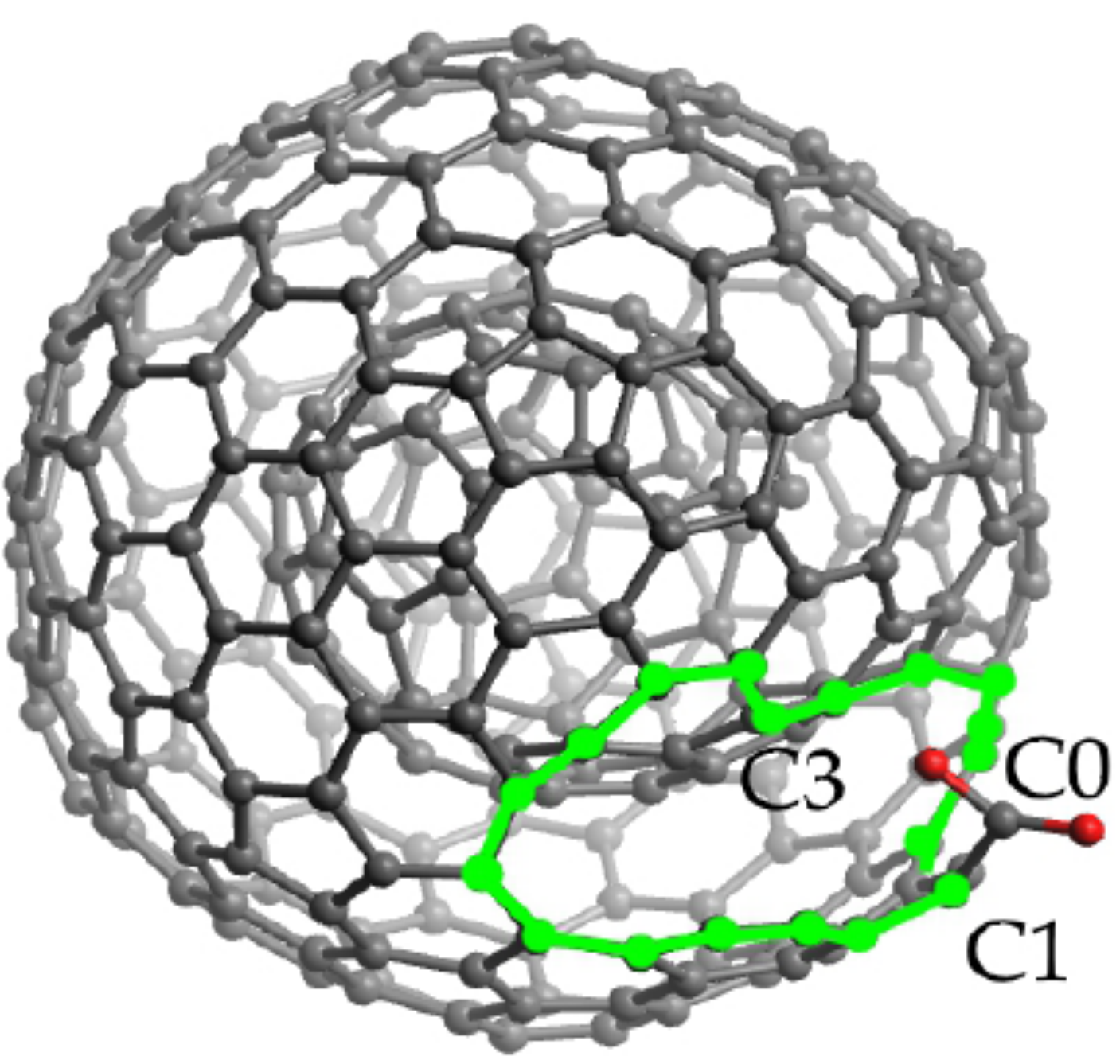}
			\caption{Optimized geometry of $CO_2$ chemisorbed on the edge of a $C_{300}$ spiroid.}
			\label{figure:spiroid}
			\end{figure}

			% Two spherical carbon allotropes have been analyzed. One of them is a $C_{60}$ fullerene enclosed in $C_{240}$ fullerene ($C_{60}@C_{240}$), or a carbon onion \cite{CarbonOnions}. In laboratory conditions, it can be synthesized via electron beam irradiation of carbon nanostructures. This process features an intermediate product in the form of carbon spiroid $C_{300}$ with the same number of atoms as the carbon onion. It is thermodynamically stable and the process of conversion between the spiroid and spheroid is reversible \cite{C300Spiroid}. $C_{300}$ in both spheroid and spiroid form is also observed in nature in interstellar medium \cite{OnionsInSpace,C300SpiroidCPMD}.

			% Geometry optimizations of the $C_{300}$ spiroid show that its end features an opening around 9.3 \si{\angstrom} wide and 3.6 \si{\angstrom} high. The opening is terminated with carbon atoms with only two saturated bonds. This means that there are two likely mechanisms of adsorption. The first one -- direct chemisorption on the spiroid termination -- saturates the dangling bonds on the edge. The second mechanism is physisorption inside the spiroid, which is facilitated by the fact that the opening at the end is large.

			Two other carbon allotropes have been analyzed: we have compared the multi-shell fullerene $C_{60}@C_{240}$ ($C_{60}$ fullerene enclosed in $C_{240}$ fullerene) with the carbon onion containing also 300 carbon atoms but consisting of two closed fullerene shells. Actually, both forms of $C_{300}$, spiroid and spheroid ones, are observed in interstellar medium \cite{OnionsInSpace,C300SpiroidCPMD}.

			Geometry optimization of the $C_{300}$ spiroid shows that its end features an opening around 9.3 \si{\angstrom} wide and 3.6 \si{\angstrom} high, as illustrated in \Cref{figure:spiroid}. The opening is terminated with carbon atoms with only two saturated bonds. Our calculations show, that the $CO_2$ can be just chemisorbed on the edge of spiroid’s orifice. There is also the possibility of physisorption of $CO_2$ inside the spiroid, which constitutes the second effective mechanism of $CO_2$ capture by spiroid $C_{300}$, which apparently is facilitated by the large opening in this structure. We would like to mention that the spheroid forms of $C_{300}$ are rather inefficient in $CO_2$ capture.

		\subsubsection{Minerals}
			\begin{figure}
			\centering
			\includegraphics[width=.45\linewidth,height=.3\linewidth,keepaspectratio]{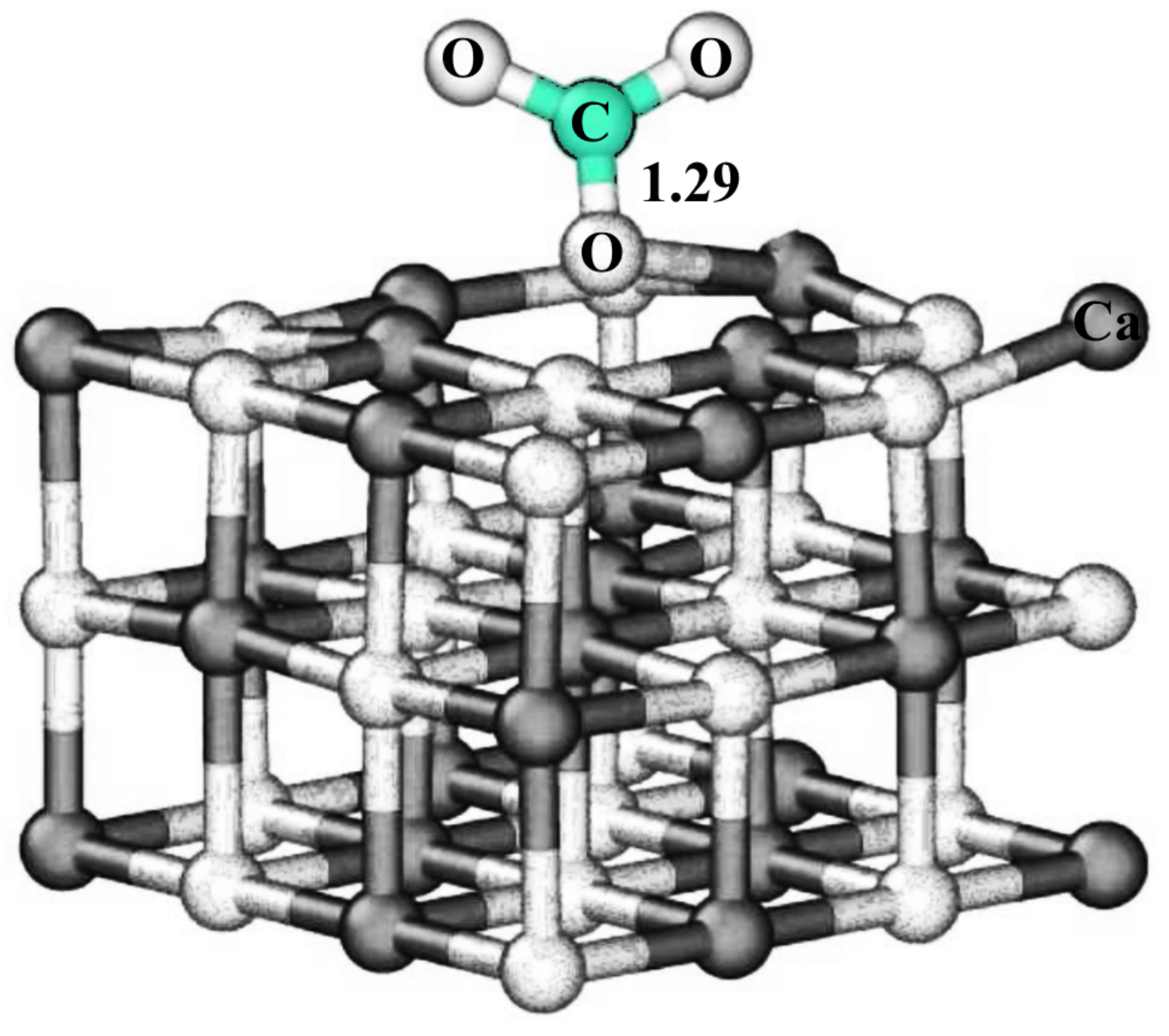}
			\includegraphics[width=.45\linewidth,height=.3\linewidth,keepaspectratio]{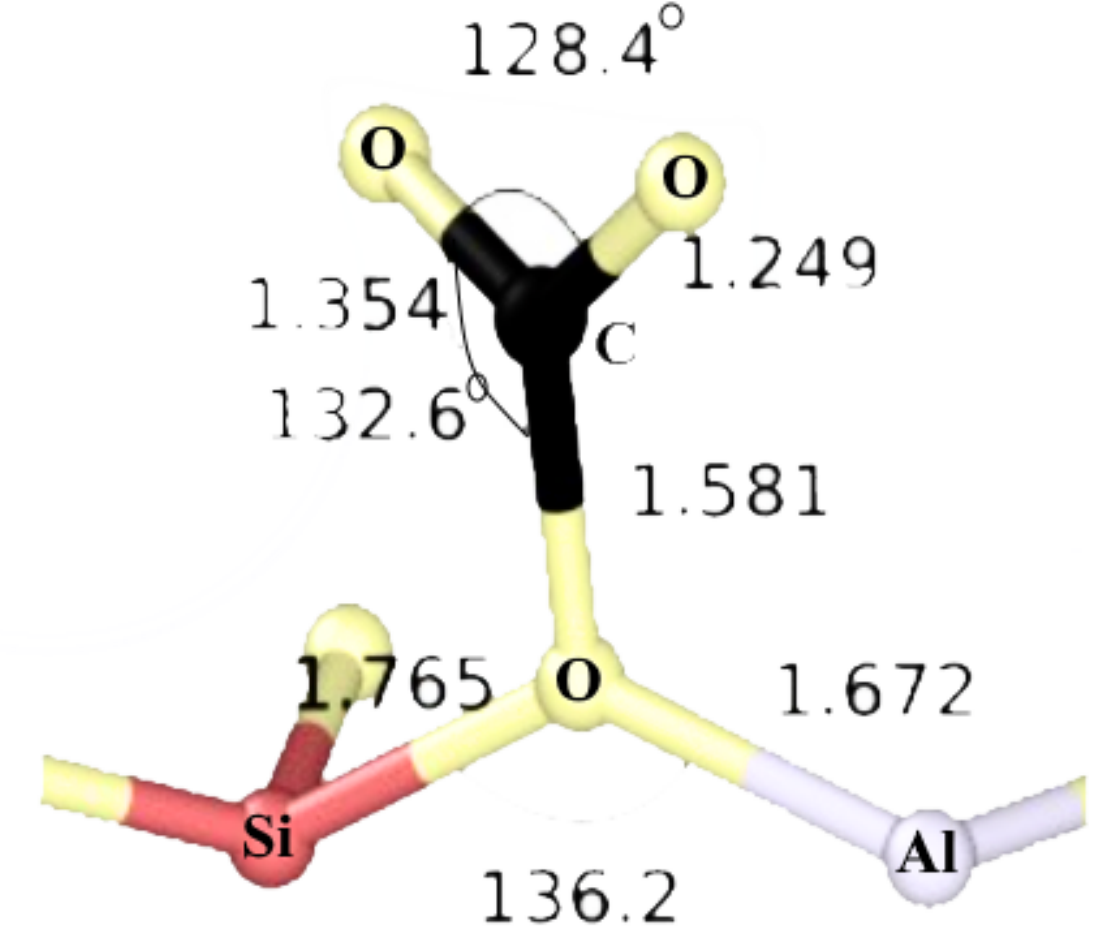}
			\caption{Optimized geometry of $CO_2$ chemisorbed on cubic MgO (left) and illite oxygen site (right).}
			\label{figure:minerals}
			\end{figure}

			We have performed AIMD simulations at temperatures up to 1200 \si{\kelvin} to determine which processes occur on which mineral surface. We found calcite to be inert to $CO_2$ adsorption up to very high temperatures. Dissociative chemisorption is preferred \cite{APPAAS} according to the equation $CO_2(gas) \rightarrow CO(ads) + O(gas)$. However, due to large energy of C-O bond breaking, it starts to occur above 1200 \si{\kelvin}. For two other minerals -- periclase and illite -- direct chemisorption is possible at low temperatures. In both cases bonding occurs via the carbon atom in $CO_2$ and oxygen atom at the surface of the mineral. The arrangements of atoms around the adsorbed $CO_2$ molecule on the MgO and illite are depicted in \Cref{figure:minerals}.

\section{Conclusions}
\label{section:Conclusions}	
	In this work, we have performed the DFT and AIMD calculations that shed light onto physicochemical adsorption mechanisms of $CO_2$ on several forms of organic and mineral materials. The highest affinity (adsorption energy) of $CO_2$ was found on the defected graphene surface. This is due to the covalent bonds formed with C and O atoms in $CO_2$. Mineral surfaces exhibited similar adsorption energies. Among them, $CO_2$ had the strongest affinity to the calcite surface. However, this reaction only occurs at extremely high temperatures starting at 1200 \si{\kelvin}. The weakest affinity is exhibited at the $C_{300}$ spiroid opening edge. The observed considerable scatter of the magnitude of the $CO_2$ affinity to the various carbon structures clearly demonstrates that the surface chemistry of carbon is very much dependent on the structural details and that the heterogeneity of organic matter can play an important role in adsorption properties of shale samples.

%% References

%The citation must be used in following style: \cite{article-minimal} \cite{article-full} \cite{article-crossref} \cite{whole-journal}.

\bibliography{EGYPRO_mwlazlo}
\bibliographystyle{elsarticle-num}
\end{document}